# Reconstructing Self Organizing Maps as Spider Graphs for better visual interpretation of large unstructured datasets

Aaditya Prakash, Infosys Limited

aaadityaprakash@gmail.com

**Abstract**--Self-Organizing Maps (SOM) are popular unsupervised artificial neural network used to reduce dimensions and visualize data. Visual interpretation from Self-Organizing Maps (SOM) has been limited due to grid approach of data representation, which makes inter-scenario analysis impossible. The paper proposes a new way to structure SOM. This model reconstructs SOM to show strength between variables as the threads of a cobweb and illuminate inter-scenario analysis. While Radar Graphs are very crude representation of spider web, this model uses more lively and realistic cobweb representation to take into account the difference in strength and length of threads. This model allows for visualization of highly unstructured dataset with large number of dimensions, common in Bigdata sources.

## INTRODUCTION

Exponential growth of data capturing devices has led to an explosion of data available. Unfortunately not all data available is in the database friendly format. Data which cannot be easily categorized, classified or imported into database are termed unstructured data. Unstructured data is ubiquitous and is assumed to be around 80% of all data generated[1]. While tremendous advancements have taken place for analyzing, mining and visualizing structured data, the field of unstructured data especially unstructured Big Data is still in nascent stage.

Lack of recognizable structure and huge size makes it very challenging to work with Unstructured Large Datasets. Classical visualization methods limit the amount of information presented and are asymptotically slow with rising dimensions of the data. We present here a model to mitigate these problems and allow efficient and vast visualization of large unstructured datasets.

A novel approach in unsupervised machine learning is Self-Organizing Maps (SOM). Along with classification, SOMs have added benefit of dimensionality reduction. SOMs are also used for visualizing multidimensional data into 2D planar diffusion map. This achieves data reduction thus enabling visualization of large datasets. Present models used to visualize SOM maps lack any deductive ability that may be defeating the power of SOM. We introduce better restructuring of SOM trained data for more meaningful interpretation of very large data sets.

Taking inspiration from the nature, we model the large unstructured dataset into spider cobweb type graphs. This has the benefit of allowing multivariate analysis as different variables can be presented into one spider graph and their inter-variable relations can be projected, which cannot be done with classical SOM maps.

## UNSTRUCTURED DATA

Unstructured Data come in different formats and sizes. Broadly, the textual data, sound, video, images, webpages, logs, emails etc. are categorized into unstructured data. In some cases, even a bundle of numeric data could be collectively unstructured for e.g. health records of a patient. While a table of Cholesterol level of all the patients is more structured, all the biostats of a single patient is largely unstructured.

### Challenges and Potential

Unstructured data could be of any form and could contain any number of independent variables. Labeling as is done in machine learning is only possible with data where information of variable such as size, length, dependency, precision etc. is known. Even extraction of the underlying information in a cluster of unstructured data is very challenging because it is not known on what is to be extracted[2]. The potential of hidden analytics within the unstructured large datasets could be a valuable asset to any business or research entity. Consider the case of Enron emails *(collected and prepared by CALO project)*. Emails are primarily unstructured, mostly because people often reply above the last email even when the new email's content and purpose might be different. Therefore most organizations do not analyze emails or logs but several researchers analyzed the Enron emails and their results show that lot of predictive and analytical information could be obtained from the same[3][4][5].

## SELF ORGANIZING MAPS

Ability to harness the increased computing power has been a great boon to business. From traditional business analytics to machine learning, the knowledge we get from data is invaluable. With computing forecasted to get faster, may be quantum computing someday, it promises greater role for the data. While there has been a lot of effort to bring some structure into unstructured data[6], the cost of doing so has been the hindrance. With larger datasets it is even a greater problem as it entails more randomness and unpredictability in the data.

Self-Organizing Maps (SOM) are class of Artificial Neural Networks proposed by Teuvo Kohonen[7] that transform the input dataset into two dimensional lattice, also called

Kohonen Map.

**Structure**

All the points of the input layer are mapped onto two dimensional lattice, called as Kohonen Network. Each point in the Kohonen Network is potentially a Neuron.

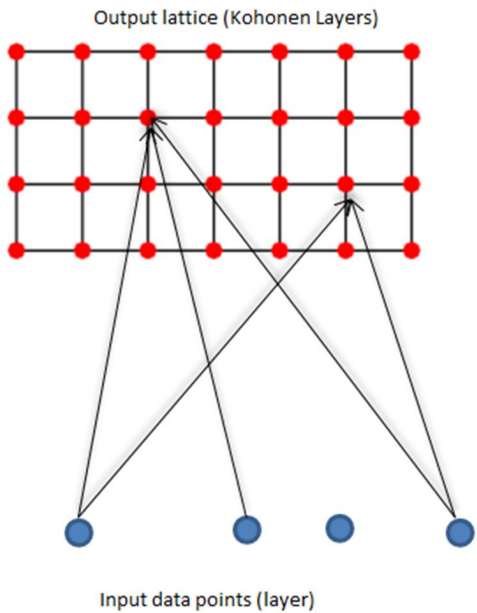

Figure 1: Kohonen Network (Source: Infosys Research)

**Competition of Neurons**

Once the Kohonen Network is completed the neurons of the network compete according to the weights assigned from the input layer. Function used to declare the winning neuron is the simple Euclidean distance of the input point and its corresponding weight for each of the neuron. The function called as discriminant function is represented as,

$$d_j(\mathbf{x}) = \sum_{i=1}^{D}(x_i - w_{ji})^2$$

where, x = point on Input Layer

w = weight of the input point (x)

i = all the input points

j = all the neurons on the lattice

d = Euclidean distance

Simply put, the winning neuron is the one whose weight is closest (distance in lattice) to the input layer. This process effectively discretizes the output layer.

**Cooperation of Neighboring Neurons**

Once the winning neuron is found, the topological structure can be determined. Similar to the behavior in human brain cells (neurons), the winning neuron also excites its neighbor. Thus the topological structure is determined by the cooperative weights of the winning neuron and its neighbor.

**Self-Organization**

The process of selecting winning neurons and formation of topological structure is adaptive. The process runs multiple times to converge on the best mapping of the given input layer. SOM is better than other clustering algorithms in that it requires very few repetitions to get to a stable structure.

**Parallel SOM for large datasets**

Among all classifying machine learning algorithms, convergence speed of the SOM

has been found to be the fastest[8]. This implies that for large data sets SOM is the best viable model.

Since the formation of topological structuring is independent of the input points it can easily be parallelized. Carpenter et al., have demonstrated the ability of SOM to work under massively parallel processing[9]. Kohonen himself has shown that even where the input data may not be in vectorial form, as found in some unstructured data, large scale SOM can be run nonetheless[10].

**SOM PLOTS**

SOM plots are a two dimensional representation of the topological structure obtained after training the neural nets for given number of repetitions and with given radius. The SOM can be visualized as a complete 2-D topological structure [Fig. 2].

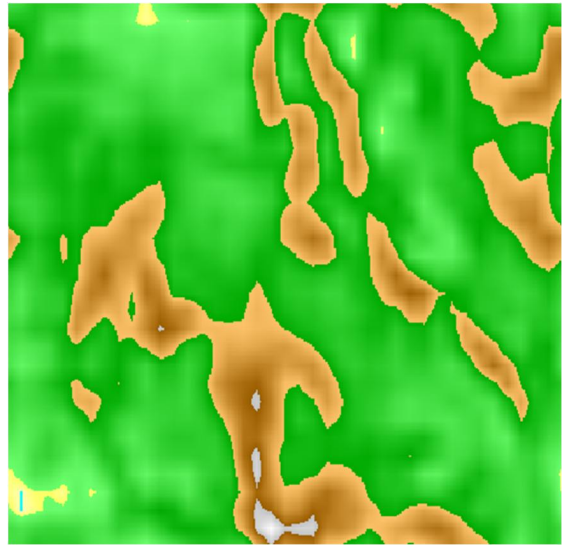

Figure 2: SOM Visualization using Rapidminer *(AGPL Open Source)* (Source: Infosys Research)

Figure 2, shows the overall topological structure obtained after dimensionality reduction of multi-variate dataset. While the graph above may be useful for outlier detection or general categorization it isn't very useful in analysis of individual variables.

Other option of visualizing SOM is to plot different variables in grid format. One can use R programming language (*GNU Open Source*) to plot the SOM results.

*Note on running example*

All the plots presented henceforth have been obtained using R programming language. Dataset used is SPAM Email Database. Database is in public domain and freely available for research at 'UCI Machine Learning Repository'. It contains 266858 word instances of 4601

SPAM emails. Emails are good example of unstructured data.

Using the public packages in R, we obtain the SOM plots.

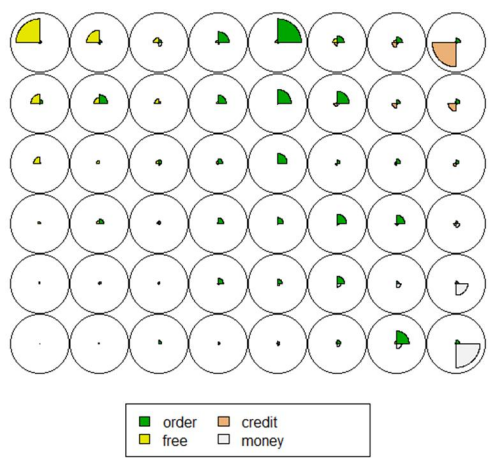

Figure 3: SOM Visualization in R using the Package 'Kohonen' (Source: Infosys Research)

Figure 3, is the plot of SOM trained result using the package 'Kohonen'[11]. This plot gives inter-variable analysis. In this case variable being 4 of one the most used words in the SPAM database viz. 'order', 'credit', 'free' and 'money'. While this plot is better than topological plot as given in Figure 2, it is still difficult to interpret the result in canonical sense.

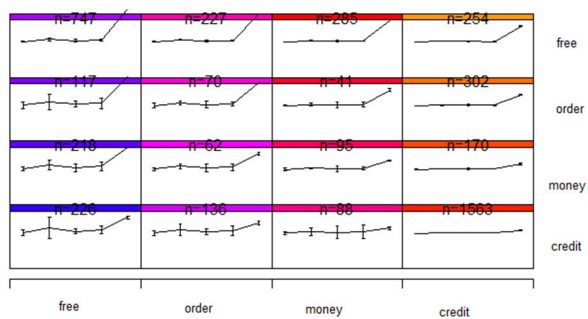

Figure 4: SOM visualization in R using the package 'SOM' (source: Infosys Research)

Figure 4, is again the SOM plot of the above given four most common words in the SPAM database but this one uses the package called 'SOM'[12]. While this plot is numerical and gives strength of inter-variate relationship it does not help in giving us the analytical picture. The information obtained is not actionable.

## SPIDER PLOTS OF SOM

As we have seen in the Figure 2, 3 and 4 the current visualization of SOM output could be improved for more analytical ability. We introduce a new method to plot SOM output especially designed for large datasets.

### Algorithm

i. Filter the results of SOM
ii. Make a polygon with as many sides as the variables in input.
iii. Make the radius of the polygon to be the maximum of the value in the dataset.
iv. Draw the grid for the polygon
v. Make segments inside the polygon if the strength of the two variables inside the segment is greater than the specified threshold.
vi. Loop Step v. for every variable against every other variable

vii. Color the segments based on the frequency of variable.
viii. Color the line segments based on the threshold of each variable pair plotted.

**Plots**

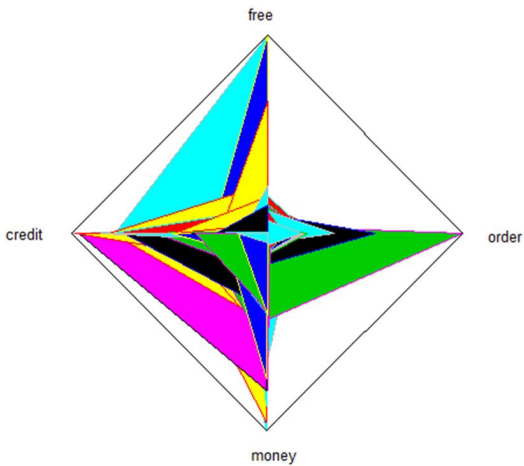

Figure 5: SOM visualization in R using the above Algorithm: Showing Segments i.e. Inter-variable Dependency (Source: Infosys Research)

As we can see, this plot is more meaningful than the SOM visualization plots obtained before. From the figure we can easily deduce that the words '*free*' and '*order*' do not have similar relation as '*credit*' and '*money*'. Understandably so, because if a Spam email is selling something, it will probably have the words '*order*' and conversely if it is advertising any product or software for '*free*' download then it wouldn't have the words '*order*' in it. High relationship between '*credit*' and '*money*' signifies Spam emails advertising for better 'Credit Score' programs and other marketing traps.

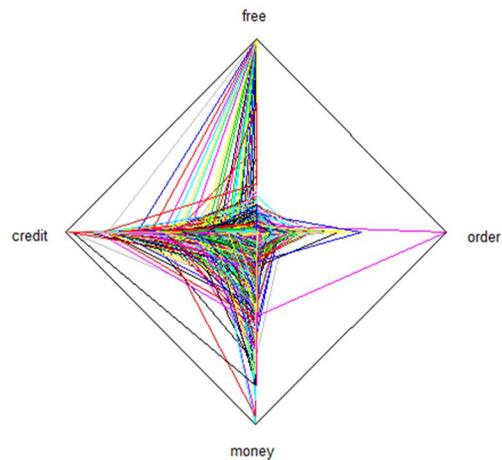

Figure 6: SOM visualization in R using above Algorithm: Showing Threads, i.e Inter-variable Strength) (Source: Infosys Research)

Figure 6 shows the relationship of each variable, in this case four popular recurring words in the Spam database. The number of threads between one variable to another shows the probability of second variable given the first variable. Several threads between '*free*' and '*credit*' suggests that Spam emails offering 'free credit' (disguised in other forms by fees or deferred interests) are among the most popular.

Using these Spider plots we can analyze several variables at once. This may cause the graph to be messy but sometimes we need to see the complete picture in order to make canonical decisions about the dataset.

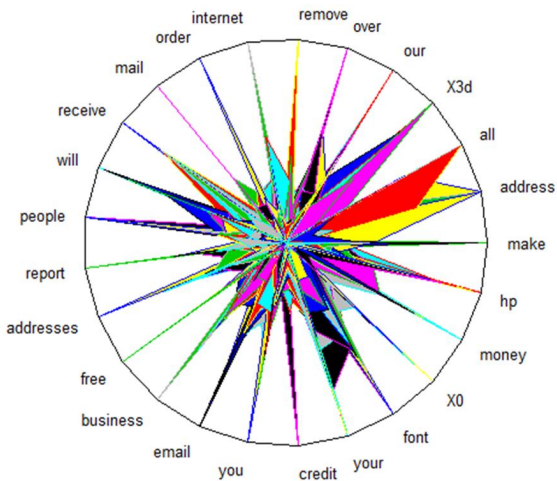

Figure 7: Spider Plot showing 25 Sampled words from the Spam Database. (Source: Infosys Research)

From the Figure 7 we can see that even though the figure shows 25 variables it is not as cluttered as a Scatter Plot or Bar chart would be if plotted with 25 variables.

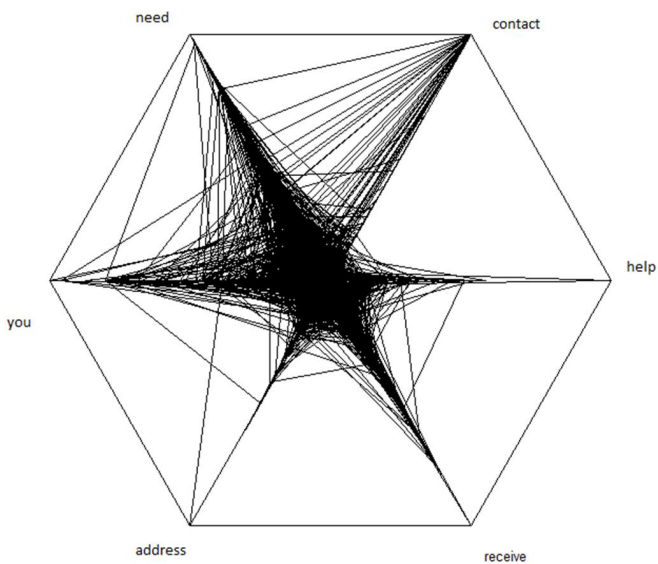

Figure 8: Uncolored Representation of threads in six variables. (Source: Infosys Research)

Figure 8 shows the different levels of strength between different variables. While '*contact*' variable is strong with '*need*' but not enough with '*help*' it is no surprise that '*you*' and '*need*' are strong. Here the idea was only to present the visualization technique and not the analysis of Spam dataset. For more analysis on Spam filtering and Spam analysis one may refer to several independent works on the same[13][14].

## CONCLUSION

While unstructured data is abundant, free and hidden with information the tools of analyzing the same are still nascent and cost of converting them to structured form is very high. Machine learning is used to classify unstructured data but comes with issues of speed and space constraints. SOM are the fastest machine learning algorithm but their visualization powers are limited. We have presented a naturally intuitive method to visualize SOM outputs which facilitates multi-variable analysis and is also highly scalable.